\documentclass[reprint,amsmath,amssymb,aps,prd,groupedaddress,showpacs,nofootinbib]{revtex4-2}

\usepackage{color}

\usepackage{bm}% bold math
\usepackage{graphicx}
\usepackage{dcolumn}% Align table columns on decimal point
\newcommand{\be}{\begin{equation}}
\newcommand{\ee}{\end{equation}}
\newcommand{\bea}{\begin{eqnarray}}
\newcommand{\eea}{\end{eqnarray}}

\newcommand{\bfa}{\mbox{\boldmath $a$}}
\newcommand{\bfr}{\mbox{\boldmath $r$}}
\newcommand{\bfx}{\mbox{\boldmath $x$}}

\newcommand{\dbss}[1]{_{\scriptstyle #1}}
\newcommand{\linda}[1]{\vphantom{A^{B}} #1}

\newcommand{\mbss}[1]{_{\mbox{\scriptsize #1}}}

\newcommand{\mbsu}[1]{\mbox{\scriptsize #1}}

\newcommand{\sscs}{\scriptscriptstyle}

\newcommand{\ve}{\varepsilon}

\newcommand{\vphi}{\varphi}

\newcommand{\etld}{\tilde{e}}
\newcommand{\gtld}{\tilde{g}}
\newcommand{\ntld}{\tilde{n}}

\newcommand{\vpa}{\vphantom{a}}
\newcommand{\vps}{\vphantom{S}}

\newcommand{\vphu}{\vphantom{*}}

\newcommand{\mft}{\mathfrak{T}}
\newcommand{\mfl}{\mathfrak{L}}

\newcommand{\norp}{n^{\prime}_{\scriptscriptstyle{+}}}
\newcommand{\nor}{n^{\vphantom{\prime}}_{\scriptscriptstyle{+}}}
\newcommand{\sssp}{\scriptscriptstyle{+}}
\begin{document}

\title{Antigravity mechanism in the theory of dual relativity}

\author{V. I. Tselyaev}
\affiliation{St. Petersburg State University, St. Petersburg, 199034, Russia}
\email{tselyaev@mail.ru}
\date{\today}

\begin{abstract}
In the paper, one of the physical consequences of the recently developed
theory of dual relativity (TDR) is considered.
The general framework of TDR is described and some results previously obtained
within this theory are summarized.
The total action functional of TDR includes the action functionals of matter fields
of two kinds: ordinary and dual.
Based on the general equations of the theory, formulas are derived for
the effective action functional of a system of point-like massive particles
belonging to both kinds of matter, in the Newtonian limit. This functional
includes an interaction term, which has the form of the gravitational interaction
energy in Newtonian mechanics.
It is shown that this energy is positive in the case of interaction between particles
of ordinary and dual matter.
This result indicates that this interaction has antigravitational nature.
\end{abstract}

\maketitle

%sec.1
\section{Introduction}

The concept of antigravity has a very long and rich history,
see, for example, a review of related ideas and works in Ref.~\cite{Nieto91}.
In the conventional sense, the term "antigravity" is associated with the concept of
a long-range repulsive interaction between massive bodies,
having a common origin with ordinary gravity.
It is worth noting that this term is sometimes used in a wider sense,
in particular, when describing the effects that arise when taking into account
the contribution of dark energy
into the equations of modern cosmological models \cite{Dolgov08,Blinnikov19}.
The possibility of antigravitational repulsion due to the geometric properties
of space-time, and/or the properties of the matter contained within it, has been
considered in many works, including modern ones, see
\cite{Bondi57,Ellis73,Bonnor89,Hossenfelder06,Hossenfelder08,
Hohmann09,Hohmann10,Hohmann14,Arbuzova14,Doroshkevich08,Novikov18a,Novikov18b,Farnes18}.
Within the framework of classical mechanics, such possibility can be formulated
by means of splitting the masses into inertial and gravitational.
Inertial mass is a kinematic object in Newton's second law.
The quantity $\sqrt{G^{\vphu}_{\mbsu{N}}} m^{\mbsu{(g)}}$, in which
$G^{\vphu}_{\mbsu{N}}$ is Newton's gravitational constant,
$m^{\mbsu{(g)}}$ is the gravitational mass, can be considered as a charge
entering the formula for the energy of gravitational interaction between two bodies
separated by a distance $r^{\vphu}_{12}$:
\be
V^{\vphu}_{12} = - G^{\vphu}_{\mbsu{N}}\,m^{\mbsu{(g)}}_1 m^{\mbsu{(g)}}_2/r^{\vphu}_{12}\,.
\label{def:v12}
\ee
The relationships between the inertial and gravitational masses of bodies
and various combinations of the signs of these masses have been analyzed in a number of works,
see, e.g., Refs.~\cite{Bondi57,Bonnor89,Hossenfelder06,Hossenfelder08,Hohmann09,Farnes18}.
In particular, if all inertial masses are positive, there are two options:
(i) both gravitational masses in (\ref{def:v12}) have the same sign,
then the interaction energy is negative,
and this interaction has the character of ordinary gravity;
(ii) the signs of $m^{\mbsu{(g)}}_1$ and $m^{\mbsu{(g)}}_2$ are opposite,
then the interaction energy is positive,
and this interaction has antigravitational character.

The simple line of reasoning presented above is, however, purely formal.
It is well known that the equivalence principle (EP)
underlying the general theory of relativity (GR)
asserts the equality of the inertial and gravitational masses of all particles of
the observable matter \cite{Landau73,Weinberg72,Ryden17}.
The possibility of the existence of matter whose particles have negative gravitational masses
needs therefore justification.
Nevertheless, although such antigravitating matter is currently only an element
of some theoretical schemes, its existence could, in principle, serve as one of the
reasons for the formation of cosmic voids, which have long been the objects of astronomical
observations and occupy the bulk of the visible Universe,
see Refs.~\cite{Ryden17,Tully08,Weygaert11}.

The aim of the present paper is to analyze antigravitational effects arising within
the framework of the recently developed theory of dual relativity (TDR) \cite{TDR}.
The possibility of such effects in the multimetric theories of gravity
was considered in Refs. \cite{Hossenfelder06,Hossenfelder08,Hohmann09,Hohmann10,Hohmann14}.
In the TDR, the analogous possibility was already mentioned in Ref.~\cite{TDR}.
It is connected with the facts that in this theory, which is generally tri-metric one,
the existence of two kinds of matter,
ordinary and dual, is assumed, and that the energy-momentum tensors of these kinds of matter
enter into the equation of motion for the gravitational field with opposite signs.
Here, antigravitational effects in the TDR will be investigated by passing
to the Newtonian limit in the general equations of this theory.

The paper is organized as follows.
The general scheme of the TDR and some results previously obtained
within its framework are briefly described in Sections \ref{sec:2} and \ref{sec:3}.
The action functionals for a system of point-like massive particles
in the Newtonian limit are derived in Sections \ref{sec:4} and \ref{sec:5}.
The effective action functional for this system and consequences following from
the structure of this functional are discussed in Sec.~\ref{sec:6}.
Conclusions are given in the last section.

%sec.2
\section{General scheme of the TDR
\label{sec:2}}

The physical space-time in the TDR is a 4D manifold ${\cal{M}}_{\linda{4}}$,
equipped with a background (nondynamical) flat metric $\gamma^{\vphu}_{\mu\nu}$ with signature $-2$.
The Greek letters $\mu,\nu,\ldots \in \{1,2,3,4\}$ here and below refer to the holonomic indices
labelling the vectors and tensors on this manifold and its coordinates $x^{\mu}$.
The lowercase Latin letters $\,a, b, \ldots \in \{1,2,3,4\}$ refer
to the anholonomic indices of vectors and tensors in the flat 4D Minkowski space $M(3,1)$
with the metric $\eta^{\vphu}_{ab} = \mbox{diag}\,\{-1,-1,-1,+1\}$.
The main geometric variables of the TDR are two vierbeins  (tetrads or local frames)
$e^a_{\mu}$ and $\etld^a_{\mu}$ and the spin connection $\omega^{ab}_{\mu}$.
The vierbeins determine two metrics
\be
g_{\dbss{\mu\nu}} = e^a_{\mu}\,\eta^{\vphu}_{ab} e^b_{\nu}
\quad\mbox{and}\quad
\gtld_{\dbss{\mu\nu}} = \etld^a_{\mu}\,\eta^{\vphu}_{ab} \etld^b_{\nu},
\label{ggtstg}
\ee
and it is assumed that they are connected with each other by the duality condition
\be
e^a_{\mu}\,\eta^{\vphu}_{ab} \etld^b_{\nu} = \gamma_{\dbss{\mu\nu}}\,.
\label{dualc}
\ee
This condition
(i) leaves only one of the two vierbeins as an independent variable,
that reduces the number of geometric variables of the TDR to the number
of variables of GR;
(ii) introduces in the theory the total energy-momentum tensor density
of the system ``matter plus gravitation'' $\,\mft^{(\mu\nu)}$
which is not equal to zero identically and for which the conservation law holds
\be
\partial_{\dbss{\mu}}\bigl(\,\mft^{(\mu\nu)}\gamma_{\dbss{\nu\lambda}}\,\bigr)
= \frac{1}{2}\,\mft^{(\mu\nu)}\,\partial_{\dbss{\lambda}}\gamma_{\dbss{\mu\nu}}.
\label{taucons}
\ee
The quantity $\mft^{(\mu\nu)}$ is the symmetric part of the tensor density
$\mft^{\mu\nu}$, playing the role of a Lagrange multiplier, ensuring the fulfillment
of the duality condition.

Initial gravitational action functional of the TDR ($S^{\vphu}_{\mbsu{g}}$)
consists of a kinetic term and an interaction term.
The kinetic term is taken in the form of the Utiyama-Kibble action
\cite{Utiyama56,Kibble61},
and depends only on the spin connection $\omega^{ab}_{\mu}$, its first derivatives,
and the mixed vierbein
\be
e^{(\zeta)a}_{\mu} = e^a_{\mu} + \zeta\,\etld^a_{\mu},
\label{def:mixvb}
\ee
where $\zeta$ is a mixing parameter.
The interaction term depends only on the vierbeins $e^a_{\mu}$ and $\etld^a_{\mu}$,
and includes the graviton mass $m^{\vpa}_{\mbsu{g}}$, the square of which
enters the common factor in this term.
Overall, $S^{\vphu}_{\mbsu{g}}$ is a {\it polynomial} functional of the fields
$e^a_{\mu}$, $\etld^a_{\mu}$, $\omega^{ab}_{\mu}$, and the first derivatives
of $\omega^{ab}_{\mu}$.

It is assumed that the spin connection $\omega^{ab}_{\mu}$ enters only into the
functional $S^{\vphu}_{\mbsu{g}}$, and therefore can be determined from the condition
\be
\delta S^{\vphu}_{\mbsu{g}}/\delta \omega^{ab}_{\mu} = 0\,.
\label{sconeq}
\ee
Solution of Eq.~(\ref{sconeq}) with respect to $\omega^{ab}_{\mu}$ gives
an expression for the spin connection via the mixed vierbein $e^{(\zeta)a}_{\mu}$.
On the other hand, the vierbein $\etld^a_{\mu}$ can be expressed via $e^a_{\mu}$
with the help of the relation (\ref{dualc}). By substituting these expressions
back into the initial formula for $S^{\vphu}_{\mbsu{g}}$, one obtains the reduced form
$\bar{S}^{\vphu}_{\mbsu{g}}$ of the functional $S^{\vphu}_{\mbsu{g}}$. It reads
\be
\bar{S}^{\vphu}_{\mbsu{g}} = \bar{S}^{\vphu}_{\mbsu{g}}[g,\gamma] = \int d^{4}x\,
\bigl(\,\bar{\mfl}^{\,\mbsu{kin}}_{\mbsu{g}} + \bar{\mfl}^{\,\mbsu{int}}_{\mbsu{g}}\,\bigr)\,,
\label{actgred}
\ee
where
\begin{subequations}
\label{def:lg}
\bea
\hspace{-2em}
2c\kappa_{\mbsu{g}}\,\bar{\mfl}^{\,\mbsu{kin}}_{\mbsu{g}}
&=& - \sqrt{-\det (g^{(\zeta)})}\,R^{(\zeta)},
\label{def:lgkin}\\
\hspace{-2em}
\frac{c \kappa_{\mbsu{g}}}{ \lambda_{\mbsu{g}}}\,
\bar{\mfl}^{\,\mbsu{int}}_{\mbsu{g}} &=&
\sqrt{-\det (g)}\,\bigl(\,3f^{\mu}_{\mu}
+ f^{\mu}_{\nu}f^{\nu}_{\mu} - f^{\mu}_{\mu}f^{\nu}_{\nu}\,\bigr)
\nonumber\\
\hspace{-2em}
&+&
\sqrt{-\det (\gtld)}\,\bigl(\,3\tilde{f}^{\mu}_{\mu}
+ \tilde{f}^{\mu}_{\nu}\tilde{f}^{\nu}_{\mu}
- \tilde{f}^{\mu}_{\mu}\tilde{f}^{\nu}_{\nu}\,\bigr),
\label{def:lgint}
\eea
\end{subequations}
$\kappa_{\mbsu{g}} = 8\pi G_{\mbsu{N}}/c^4$,
$\lambda_{\mbsu{g}} = (m^{\vpa}_{\mbsu{g}}c/\hbar)^2/8$,
$m^{\vpa}_{\mbsu{g}}$ denotes the graviton mass at $\zeta = 0$,
\be
g^{(\zeta)}_{\mu\nu} = \,e^{(\zeta)a}_{\mu}\,\eta^{\vphu}_{ab}\,e^{(\zeta)b}_{\nu}\,
= \,g_{\dbss{\mu\nu}} + 2\zeta \gamma_{\dbss{\mu\nu}}
+ \zeta^2 \gtld_{\dbss{\mu\nu}}\,,
\label{def:gzet}
\ee
$R^{(\zeta)}$ is the Ricci scalar constructed from the metric tensor
$g^{(\zeta)}_{\mu\nu}$,
\begin{subequations}
\label{def:ftfmn}
\bea
f^{\mu}_{\nu} &=& g^{\mu\lambda}\gamma_{\dbss{\lambda\nu}}
= \gamma^{\mu\lambda}\gtld_{\dbss{\lambda\nu}}\,,
\label{def:fmn}\\
\tilde{f}^{\mu}_{\nu} &=& \gamma^{\mu\lambda}g_{\dbss{\lambda\nu}}
= \gtld^{\mu\lambda}\gamma_{\dbss{\lambda\nu}}\,.
\label{def:tfmn}
\eea
\end{subequations}
The metric $\gtld_{\dbss{\mu\nu}}$ in these formulas is determined via
the metrics $g_{\dbss{\mu\nu}}$ and $\gamma_{\dbss{\mu\nu}}$ by the relation
\be
\gtld_{\dbss{\mu\nu}} =
\gamma_{\dbss{\mu\lambda}} g^{\lambda\kappa} \gamma_{\dbss{\kappa\nu}}
\label{ggtrel}
\ee
following from the duality condition. Thus,
the functional $\bar{S}^{\vphu}_{\mbsu{g}}$ depends only on the metrics
$g_{\dbss{\mu\nu}}$ and $\gamma_{\dbss{\mu\nu}}$ and their derivatives,
but already has no polynomial form.
Compared to the gravitational functional of GR, this functional contains
two additional constants:
the graviton mass $m^{\vpa}_{\mbsu{g}}$ and the mixing parameter $\zeta$.
When these additional constants are zero, the gravitational action functionals
of the TDR and GR effectively coincide.

The total action functional of the TDR includes the action functionals of the fields
of ordinary and dual matter.
It is assumed that the dynamics of ordinary matter fields is determined by
the vierbein $e^a_{\mu}$ and the metric $g_{\dbss{\mu\nu}}$,
and the dynamics of dual matter fields is determined by the vierbein $\etld^a_{\mu}$
and the metric $\gtld_{\dbss{\mu\nu}}$.
It is also assumed that the sets of the fields of ordinary and dual matter
have no common elements and therefore are coupled to each other
{\it only} by means of the relations (\ref{dualc}) and (\ref{ggtrel}).
In this case these fields are ``dark'' for antipodal observers
(in accordance with the terminology of
Ref.~\cite{Hossenfelder08}, this means that the fields of dual matter are ``dark''
for the $g$-observer, while the fields of ordinary matter are ``dark'' for
the $\tilde{g}$-observer).
However, this assumption, call it the assumption of minimal coupling,
is not mandatory and is made only to simplify some consequences of the theory.

Note that, as was pointed out in \cite{Hohmann14}, if two kinds of matter,
whose motion is determined by two different, albeit related to each other,
metrics, are introduced in the theory, the EP is generally violated.
There are, however, various formulations and versions of this principle,
see, in particular, Refs.~\cite{Ohanian77,Mannheim06,Clifton12,Casola15} in addition
to Refs. \cite{Landau73,Weinberg72,Ryden17} mentioned in the Introduction.
Without going into details, one can say that the EP establishes the local equivalence
between the gravity and inertia. But it does not determine the origin and form
of the gravitational field itself. At the microscopic level, it only constrains
admissible forms of the action functionals of matter.
This makes it possible to formulate a principle which can be called
the principle of alternative equivalences, and which, in a sense, is a specified version
of the modified weak equivalence principle introduced in Ref.~\cite{Hohmann09}.
It implies that the local equivalence
between the gravity and inertia can take place (at least for the macroscopic bodies)
either separately for the ordinary matter or separately for the dual matter,
but not for both these kinds of matter taken together.
In the TDR, this principle is valid only if the above-mentioned
assumption of minimal coupling is fulfilled, because only in this case
the action functionals of both kinds of matter can have the forms analogous to
the form of the respective functionals of the GR.
Nevertheless, in any case, all the initial action functionals of the TDR satisfy
other conditions that are usually imposed on the functionals of the GR
and its modifications.
These are the conditions of invariance under general coordinate transformations
in ${\cal{M}}_{\linda{4}}$
and under local Lorentz transformations acting on anholonomic indices
of vectors and tensors in Minkowski space $M(3,1)$.

%sec.3
\section{Cosmological solutions
\label{sec:3}}

The TDR is constructed in such a way that the successful description of the
gravitational effects on the macroscopic scales achieved within the conventional GR
could be reproduced.
However, the cosmology leaves many possibilities for developing the modified theories
of gravity, and one of these possibilities is realized in the TDR.
In the cosmological limit, $g_{\dbss{\mu\nu}}$ is assumed to be a spatially flat
metric of the form
\be
g_{\dbss{\mu\nu}} = \mbox{diag}\,\{-A^2,-A^2,-A^2,\,B^2\},
\label{def:gclim}
\ee
where $A$ is a scale factor, $B$ is a lapse function.
It is assumed that functions $A$ and $B$ depend only on time $t$.
The coordinate system is chosen in such a way that
\be
\gamma^{\vphu}_{\mu\nu} = \mbox{diag}\,\{-1,-1,-1,+1\}.
\label{def:gamm}
\ee
The equations of motion of the TDR in this limit lead to an algebraic equation
defining $B$ as a function of $A$, and to the following equation for $A(t)$
\be
\dot{A}^2 + {\mathcal U}_{\mbss{c}}(A) = 0\,,
\label{a:eom}
\ee
where $\dot{A} = dA/dt$ and $\,{\mathcal U}_{\mbss{c}}(A)$ is
an algebraic function of $A$.
This equation has the form of a first integral of the equation
of one-dimensional motion of a fictitious ``particle'' with zero ``energy'',
in which $A$ is the coordinate of the ``particle'',
and the function ${\mathcal U}_{\mbss{c}}(A)$ has the meaning of a (cosmological)
quasipotential.

There are two types of solutions to Eq.~(\ref{a:eom}):
with positive and with negative total energy density $\ve_{\mbsu{tot}}$,
which is defined as $\ve_{\mbsu{tot}} = \mbox{sgn}(B)\,\mft^{44}$, and
which enters ${\mathcal U}_{\mbss{c}}(A)$ as a parameter.
Among the solutions with $\ve_{\mbsu{tot}}>0$, there are those that correspond
to a stable Universe taken as a whole, with an equilibrium metric coinciding
with the background flat metric $\gamma^{\vphu}_{\mu\nu}$.
The existence of these solutions is a consequence of the fact that in the general case,
the equations of motion for the gravitational field in the TDR,
see Sec.~IV.C in \cite{TDR}, have the formal solution
\be
g_{\dbss{\mu\nu}} = \gamma^{\vphu}_{\mu\nu}
%g_{\dbss{\mu\nu}} = \gtld_{\dbss{\mu\nu}} = \gamma^{\vphu}_{\mu\nu},
\label{sts1}
\ee
under the condition
\be
\bigl( T_{\dbss{\mu\nu}} \bigr)_{g = \gamma} =
\bigl( \tilde{T}_{\dbss{\mu\nu}} \bigr)_{\gtld = \gamma}\,,
\label{sts2}
\ee
where $T_{\dbss{\mu\nu}}$ and $\tilde{T}_{\dbss{\mu\nu}}$ are the energy-momentum tensors
of the ordinary and dual matter, respectively.
Locally, Eq.~(\ref{sts2}) is obviously not fulfilled. But it can be fulfilled on average,
on the cosmological scales.
In the cosmological limit, in which the model of the perfect fluid for the matter is used,
Eq.~(\ref{sts2}) is equivalent to the conditions
\be
\ve^{(\mbsu{st})}_{\mbsu{m}} = \tilde{\ve}^{(\mbsu{st})}_{\mbsu{m}},
\qquad p^{(\mbsu{st})} = \tilde{p}^{(\mbsu{st})},
\label{epcond}
\ee
where $\ve^{(\mbsu{st})}_{\mbsu{m}}$ and $\tilde{\ve}^{(\mbsu{st})}_{\mbsu{m}}$
($p^{(\mbsu{st})}$ and $\tilde{p}^{(\mbsu{st})}$) are
the energy densities (pressures) of ordinary and dual matter, respectively,
taken at the stability point (\ref{sts1}).
If Eqs. (\ref{epcond}) hold, and
$\ve_{\mbsu{tot}} =
\ve^{(\mbsu{st})}_{\mbsu{m}} + \tilde{\ve}^{(\mbsu{st})}_{\mbsu{m}}$,
the necessary conditions of stability of solutions to Eq.~(\ref{a:eom}),
${\mathcal U}^{\vphu}_{\mbsu{c}}(1) = {\mathcal U}^{\prime}_{\mbsu{c}}(1) = 0$,
are satisfied, see \cite{TDR}.
But the conditions (\ref{epcond}) together with the additional stability condition,
${\mathcal U}^{\prime\prime}_{\mbsu{c}}(1) > 0$,
can be fulfilled only if both kinds of matter, ordinary and dual, exist,
and therefore the existence of dual matter is necessary for the existence
of solutions of the given type.

Solutions with $\ve_{\mbsu{tot}}<0$ describe oscillations of the scale factor
within finite limits, determined by the turning points.
The point $A=0$ is outside these limits, and thus singularity at the zero point is excluded.
This enables one to avoid the known problem of the initial singularity
\cite{Overduin98,Novello08}, which in most standard approaches based on the GR
is interpreted in terms of the Big Bang model.
However, the cosmological quasipotential ${\mathcal U}_{\mbss{c}}(A)$ in the TDR for solutions
with $\ve_{\mbsu{tot}}<0$ has a quasisingular character at the other two critical points and
in the region between them, where it tends to $-\infty$ as $\zeta \rightarrow +0$
(but remains finite for finite $\zeta > 0$).
This dependence of the quasipotential on $A$ signifies a sharp change of the scale factor
and, consequently, the density of matter at small $\zeta$: a change from the value of $A$
at one critical point to the value at another over an extremely short time interval $T_{\zeta}$.
Using the terminology adopted in modern cosmology \cite{Ryden17},
this sharp change of $A$ (by at least a factor of $\sqrt{3}$)
can be called a ``small bang'' when the scale factor increases
and a ``small crunch'' when it decreases.

\begin{figure}[]
\begin{center}
\includegraphics*[trim=3cm 15cm 0cm 2cm,clip=true,scale=0.5,angle=0]{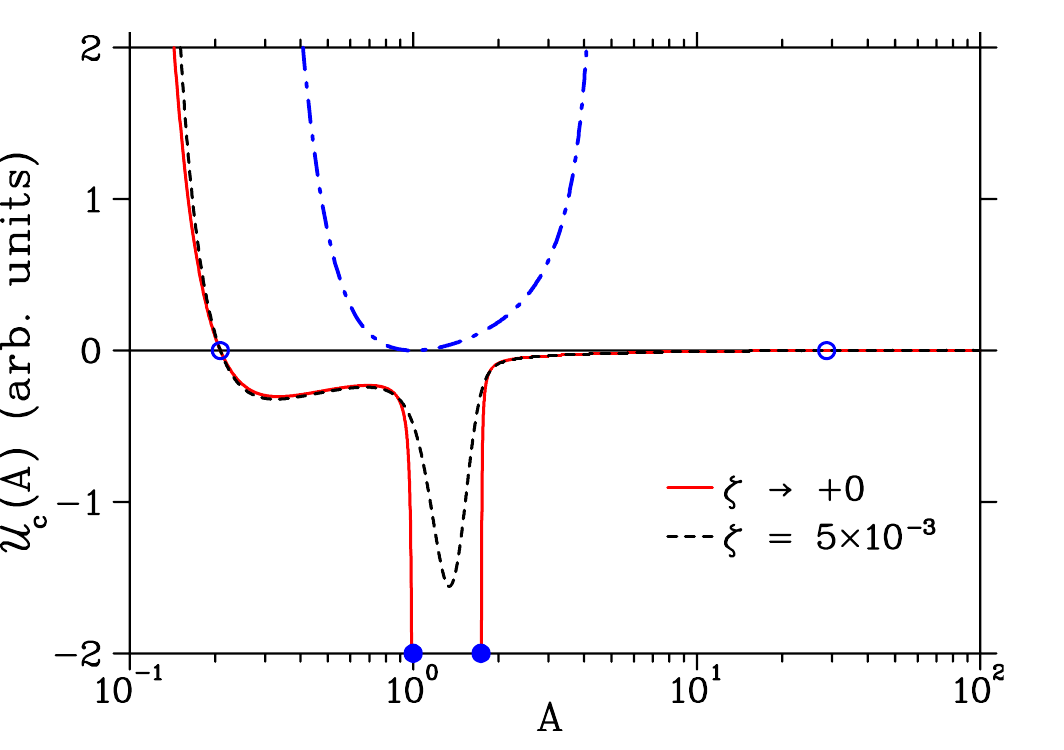}
\end{center}
\caption{\label{fig:u2}
Typical forms of the cosmological quasipotential ${\mathcal U}_{\mbss{c}}(A)$.
The red solid line corresponds to the oscillating solution with $\ve_{\mbsu{tot}}<0$
and with parameters fitted in Ref.~\cite{TDR} to the set WM57 of the $H(z)$ data
from Ref.~\cite{Farooq17}. The critical points of ${\mathcal U}_{\mbss{c}}(A)$,
$A_{\mbsu{c.L}} = 1$ and $A_{\mbsu{c.R}} = \sqrt{3}$,
are indicated by the blue filled circles on the $A$-axis.
The turning points of the oscillations are indicated by the blue open circles.
The black dashed line demonstrates the general character
of the function ${\mathcal U}_{\mbss{c}}(A)$ calculated with $\ve_{\mbsu{tot}}<0$
at not too small value of the mixing parameter $\zeta$.
The blue dashed-dotted line corresponds to the stable solution with $\ve_{\mbsu{tot}}>0$
at $\zeta \rightarrow +0$.
}
\end{figure}

The existence of two types of solutions to the TDR equations in the cosmological limit
allows one to suppose that the Universe as a whole is stable and that the oscillating
solutions describe the evolution of the scale factor in finite regions of the Universe
(domains) whose sizes can be very large.
These oscillating solutions can be used to determine the model parameters, in particular,
to determine the graviton mass $m^{\vphu}_{\mbsu{g}}$,
by fitting the calculated values of the squared Hubble parameter $H^2(A)=\dot{A}^2/A^2$
to measured data on the dependence of $H$ on the redshift parameter $z$.
The values of the TDR parameters found in this way are presented in \cite{TDR}.
For all $H(z)$ data sets used,
the $\chi^2$ value, which determines the deviation
of the calculated $H^2$ values from the measured data and which is minimized
in the fitting procedure, is smaller (in some cases significantly smaller)
in the TDR than in the $\Lambda$CDM model \cite{Peebles84}, which is the standard model
of modern cosmology.
The typical forms of the function ${\mathcal U}_{\mbss{c}}(A)$ for the solutions
with $\ve_{\mbsu{tot}}<0$ and $\ve_{\mbsu{tot}}>0$ are shown in Fig.~\ref{fig:u2}.

The obtained Compton wavelength of the graviton
$l^{\vphu}_{\mbsu{g}}=\hbar/m^{\vphu}_{\mbsu{g}}c$,
which coincides in order of magnitude with the range of gravity,
is, for different sets of $H(z)$ data, in the interval from 24.2 Gpc to 64.1 Gpc.
On average, this is an order of magnitude larger than the so-called Hubble length
$c/H^{\vphu}_0 \approx$ 4.4 Gpc $\approx 1.4 \times 10^{26}$ m,
where $H^{\vphu}_0$ is the current value of the Hubble parameter,
obtained in the $\Lambda$CDM model.

This method does not allow one to determine directly the value of the mixing parameter
introduced in the TDR due to its assumed extreme smallness. The estimate used in \cite{TDR},
$\zeta \approx m^{\vphu}_{\mbsu{g}}/m^{\vphu}_{\mbsu{P}}$,
where $m^{\vphu}_{\mbsu{P}}$ is the Planck mass, yields $\zeta \sim 10^{-62}$.
For such values of $\zeta$, the ``jump'' time
of the scale factor $T_{\zeta} \sim 10^{-27}$ s, which is much less than the typical time
of direct nuclear reactions ($\sim 10^{-22}$ s).

Other time characteristics of the scale-factor oscillations were calculated in \cite{TDR}
using parameters of the cosmological model determined in the fitting procedure.
The obtained values of the time interval $T_{\mbsu{C}}$ remaining
until the next ``small bang'' (i.e., until the left critical point $A_{\mbsu{c.L}}$ of the
quasipotential will be reached) have a large dispersion from 17 Myr to 1.8 Gyr.
This spread is explained by the fact that the value of $T_{\mbsu{C}}$ is determined by the dependence
${\mathcal U}_{\mbss{c}}(A)$ in the region of $A$ where $H(z)$ data are absent, and by the fact that
some parameters of the function ${\mathcal U}_{\mbss{c}}(A)$ are very sensitive to the details
of the observed distribution of $H(z)$, which has large uncertainties for the existing data.

The effective value of the cosmological constant $\Lambda$, one of the main elements
of many cosmological models (see, e.g.,
\cite{Dolgov08,Blinnikov19,Ryden17,Overduin98,Novello08,Peebles84,Farooq17,Weinberg89,Martin12}),
in the TDR is exactly zero.
This resolves, at least at the classical level, the problem related with this constant,
which has long been actively discussed in the literature.
The $\Lambda$-term of the action functional is usually associated with the contribution of dark energy.
In the $\Lambda$CDM model, the constant $\Lambda$ serves to explain the observed cosmic acceleration.
In the TDR, this acceleration effect is achieved at non-zero negative values
of the total energy density $\ve_{\mbsu{tot}}$ and a non-zero graviton mass in the region
where the scale factor $A$ approaches the left critical point $A_{\mbsu{c.L}}$
of the quasipotential ${\mathcal U}_{\mbss{c}}(A)$, see Fig.~\ref{fig:u2}.
Thus, {\it there is no dark energy in the TDR}.

%sec.4
\section{Matter action functional in the Newtonian limit
\label{sec:4}}

Consider a system of point-like massive particles interacting with each other
only by means of the forces determined by the properties of the metrics
($g_{\dbss{\mu\nu}}$ for ordinary matter and $\gtld_{\dbss{\mu\nu}}$ for dual matter),
which govern their motion.
The action functional for a system of such particles referring to ordinary matter
can be written as (see \cite{Weinberg72}):
\be
S^{\vphu}_{\mbsu{m}} =
- \sum_n m^{\vphu}_n c \int^{\,\infty}_{-\infty} dp
\Bigl( g_{\dbss{\mu\nu}} \frac{dx^{\mu}_n(p)}{dp} \frac{dx^{\nu}_n(p)}{dp}
\Bigr)^{1/2},
\label{def:sm}
\ee
where $n$ is the number of the particle,
$x^{\mu}_n(p)$ and $m^{\vphu}_n$ stand for its coordinate and mass, respectively
(it is assumed that $m^{\vphu}_n > 0$),
$p$ is the particle trajectory parameter, and $c$ is the speed of light.
After changing the integration variable $p \rightarrow t = x^4_n(p)/c$ for each $n$,
the formula (\ref{def:sm}) takes the form:
\bea
S^{\vphu}_{\mbsu{m}} &=& - \sum_n m^{\vphu}_n c^2
\nonumber\\
& \times & \int^{\,\infty}_{-\infty} dt
\Bigl( g_{\dbss{44}}
+ 2 g_{\dbss{4k}} \frac{\dot{x}^k_n}{c}
+ g_{\dbss{kl}} \frac{\dot{x}^k_n\dot{x}^l_n}{c^2}
\Bigr)^{1/2},
\label{def:smr}
\eea
where $k,l,\ldots \in \{1,2,3\}$ refer to the holonomic spatial indices
of the coordinates, vectors, and tensors, $\dot{x}^k_n = d x^k_n(t)/dt$.
It is assumed that the system of particles located within a domain with $\ve_{\mbsu{tot}}<0$
is considered.
The metric $g_{\dbss{\mu\nu}}$ averaged over the volume of this domain and
denoted here and in what follows as $g^{(0)}_{\dbss{\mu\nu}}$,
is defined by Eq.~(\ref{def:gclim}).
However, in this case, it is convenient to choose a different coordinate system
in which this metric has the form:
\be
g^{(0)}_{\dbss{\mu\nu}} = \mbox{diag}\,\{-1,-1,-1,+1\}.
\label{def:g0}
\ee
In this coordinate system, one has for the background metric $\gamma^{\vphu}_{\mu\nu}$,
instead of (\ref{def:gamm}):
\be
\gamma_{\dbss{\mu\nu}} = \mbox{diag}\,\{-A^{-2},-A^{-2},-A^{-2},\,B^{-2}\}.
\label{gam:ns}
\ee
Formula for the averaged metric $\gtld^{(0)}_{\dbss{\mu\nu}}$ is obtained from (\ref{ggtrel}),
(\ref{def:g0}) and (\ref{gam:ns}):
\be
\gtld^{(0)}_{\dbss{\mu\nu}} = \mbox{diag}\,\{-A^{-4},-A^{-4},-A^{-4},\,B^{-4}\}.
\label{gt0:ns}
\ee
The functions $A(t)$ and $B(t)$, determined by the TDR equations in the cosmological limit,
are taken in Eqs. (\ref{gam:ns}) and (\ref{gt0:ns}) at some fixed value $t=t_{\dbss{0}}$,
and it is assumed that they are approximately constant over a sufficiently broad interval
around this value of $t$.

Let $h_{\dbss{\mu\nu}}$ be the deviation of the metric $g_{\dbss{\mu\nu}}$
from its averaged value $g^{(0)}_{\dbss{\mu\nu}}$, i.e.
\be
g_{\dbss{\mu\nu}} = g^{(0)}_{\dbss{\mu\nu}} + h_{\dbss{\mu\nu}}.
\label{def:hmn}
\ee
It is advisable to introduce, following \cite{Landau73}, the notation
\be
h_{\dbss{44}} = \frac{2\vphi}{c^2}
\label{def:h44}
\ee
and to make change of variables
\be
h_{\dbss{4k}} =\, h_{\dbss{k4}} = \frac{2\xi_{\dbss{k}}}{c^2}\,,\quad
h_{\dbss{kl}} = \frac{2\vphi\delta_{\dbss{kl}} + 2\psi_{\dbss{kl}}}{c^2}\,.
\label{def:h4kkl}
\ee
In the Newtonian limit, which assumes the smallness of the values of $h_{\dbss{\mu\nu}}$,
the smallness of the ratios of $\dot{x}^k_n/c$, and the absence of the explicit dependence
of the components of $h_{\dbss{\mu\nu}}$ on time,
one has from (\ref{def:smr}), (\ref{def:hmn})--(\ref{def:h4kkl})
\be
S^{\mbsu{NL}}_{\mbsu{m}} = \int^{\,\infty}_{-\infty} dt\,L^{\vphu}_{\mbsu{m}}\,,
\label{def:smnl}
\ee
where
\be
L^{\vphu}_{\mbsu{m}} = \sum_n
\Bigl(\,\frac{1}{2}\,m^{\mbsu{(i)}}_n \dot{\bfx}^2_n
- m^{\mbsu{(g)}}_n \vphi(\bfx^{\vphu}_n) - m^{\mbsu{(g)}}_n\,c^2\,
\Bigr),
\label{def:lmnl}
\ee
and the notations for the inertial (i) and gravitational (g) masses of particles
of ordinary matter were introduced:
\be
m^{\mbsu{(i)}}_n = m^{\mbsu{(g)}}_n = m^{\vphu}_n.
\label{mig}
\ee
The Lagrangian $L^{\vphu}_{\mbsu{m}}$ is constrained by the terms of the first order
in $h_{\dbss{\mu\nu}}$ and zeroth order in $\dot{x}^k_n/c$.
That the quantities $m^{\mbsu{(i)}}_n$ are the inertial masses of particles
is clear directly from Eq.~(\ref{def:lmnl}) for $L^{\vphu}_{\mbsu{m}}$.
The validity of identifying the quantities $m^{\mbsu{(g)}}_n$ in (\ref{def:lmnl})
with gravitational masses will be justified below.

The initial action functional for the system of dual matter particles has a form
similar to (\ref{def:sm}):
\be
\tilde{S}^{\vphu}_{\mbsu{m}} =
- \sum_{\ntld} m^{\vphu}_{\ntld} c \int^{\,\infty}_{-\infty} dp
\Bigl( \gtld_{\dbss{\mu\nu}} \frac{dx^{\mu}_{\ntld}(p)}{dp} \frac{dx^{\nu}_{\ntld}(p)}{dp}
\Bigr)^{1/2},
\label{def:smt}
\ee
where, as in (\ref{def:sm}), $m^{\vphu}_{\ntld} > 0$.
For the metric $\gtld_{\dbss{\mu\nu}}$ entering (\ref{def:smt}) one obtains
from Eqs. (\ref{ggtrel}) and (\ref{def:hmn})
\be
\gtld^{\vps}_{\dbss{\mu\nu}} = \gtld^{(0)}_{\dbss{\mu\nu}}
- \tilde{h}^{\vps}_{\dbss{\mu\nu}},
\label{gt0h}
\ee
where up to linear terms in $h_{\dbss{\mu\nu}}$
\be
\tilde{h}^{\vps}_{\dbss{\mu\nu}} = \gamma_{\dbss{\mu\lambda}}g^{(0)\lambda\lambda'}
h_{\dbss{\lambda'\kappa'}}g^{(0)\kappa'\kappa}\gamma_{\dbss{\kappa\nu}}.
\label{def:htmn}
\ee
Here the tensor $g^{(0)\mu\nu}$ is the inverse metric to $g^{(0)}_{\mu\nu}$, i.e.
$g^{(0)\mu\lambda}g^{(0)}_{\lambda\nu} = \delta^{\mu}_{\dbss{\nu}}$,
the metrics $g^{(0)}_{\mu\nu}$, $\gamma^{\vps}_{\mu\nu}$, and $\gtld^{(0)}_{\dbss{\mu\nu}}$
are determined be Eqs. (\ref{def:g0})--(\ref{gt0:ns}).
Note that the opposite signs in front of the second terms in the right-hand sides (r.h.s.)
of Eqs. (\ref{def:hmn}) and (\ref{gt0h}) are a consequence of the relation (\ref{ggtrel}).

In analogy with the derivation scheme of Eq.~(\ref{def:lmnl}), one obtains for the
dual matter action functional (\ref{def:smt}) in the Newtonian limit
\be
\tilde{S}^{\mbsu{NL}}_{\mbsu{m}} = \int^{\,\infty}_{-\infty} dt\,
\tilde{L}^{\vphu}_{\mbsu{m}}\,,
\label{def:smtnl}
\ee
where
\be
\tilde{L}^{\vphu}_{\mbsu{m}} = \sum_{\ntld}
\Bigl(\,\frac{1}{2}\,m^{\mbsu{(i)}}_{\ntld} \dot{\bfx}^2_{\ntld}
- m^{\mbsu{(g)}}_{\ntld} \vphi(\bfx^{\vphu}_{\ntld}) - |m^{\mbsu{(g)}}_{\ntld}|\,c^2\,
\Bigr).
\label{def:lmtnl}
\ee
Here, however, in contrast to the equality between the inertial and gravitational masses
of particles of ordinary matter, one has
\be
m^{\mbsu{(i)}}_{\ntld} = \frac{B^2}{A^4}\,m^{\vphu}_{\ntld}\,,\qquad
m^{\mbsu{(g)}}_{\ntld} = - \frac{A^4}{B^4}\,m^{\mbsu{(i)}}_{\ntld}.
\label{mtig}
\ee
From Eqs. (\ref{mtig}) it is seen that for $m^{\vphu}_{\ntld} > 0$ in (\ref{def:smt}),
the inertial masses of the dual matter particles stay positive,
but their gravitational masses are negative, due to the negative sign
in front of the tensor $\tilde{h}^{\vps}_{\dbss{\mu\nu}}$ in the r.h.s. of (\ref{gt0h}).
Furthermore, in absolute value, $|m^{\mbsu{(g)}}_{\ntld}| \neq |m^{\mbsu{(i)}}_{\ntld}|$
for $A^2 \neq B^2$.

The present-day values $A^{\vphu}_0 = A(t^{\vphu}_0)$ and $B^{\vphu}_0 = B (t^{\vphu}_0)$
were determined in Ref.~\cite{TDR}, along with other model parameters,
by fitting to $H(z)$ data. The resulting values of $A^{\vphu}_0$
are close to 1 (and less than 1 for all data sets). The ratios
$A^2_0/B^2_0$ range from $4.5\times 10^2$ to $1.2\times 10^5$.

Note that the inequality $A^2/B^2 > 1$ implies the possibility of ``superluminal'' motion
of dual matter particles relative to ordinary matter particles,
i.e., motion at velocities greater than the speed of light of ordinary matter.
The situation is opposite for values of $t$ for which $A^2/B^2 < 1$.
In this case, the ordinary matter has ``superluminal'' properties in relation
to the dual matter, see \cite{TDR} for more details.

The total action functional of a system of particles of ordinary and dual matter
in the Newtonian limit can be written as
\be
S^{\mbsu{NL}}_{\mbsu{m}\sssp} =
S^{\mbsu{NL}}_{\mbsu{m}} + \tilde{S}^{\mbsu{NL}}_{\mbsu{m}} =
\int^{\,\infty}_{-\infty} dt\,L^{\vphu}_{\mbsu{m}\sssp}\,,
\label{def:smtotnl}
\ee
where
\be
L^{\vphu}_{\mbsu{m}\sssp} = T - V^{\vphu}_0 - M^{\vphu}_{\mbsu{g}\sssp}\,c^2,
\label{def:lmtotnl}
\ee
\be
T = \frac{1}{2} \sum_{\nor} m^{\mbsu{(i)}}_{\nor} \dot{\bfx}^2_{\nor}\,,
\quad
V^{\vphu}_0 = \sum_{\nor} m^{\mbsu{(g)}}_{\nor} \vphi(\bfx^{\vphu}_{\nor})\,,
\label{def:tvnl}
\ee
\be
M^{\vphu}_{\mbsu{g}\sssp} = \sum_{\nor} |m^{\mbsu{(g)}}_{\nor}|\,,
\label{def:mtot}
\ee
and the notation $\nor$ was introduced for indices belonging to the set
$\{\nor\} = \{n\} \bigcup\,\{\ntld\}$, which includes indices of particles
of both ordinary and dual matter.

%sec.5
\section{Gravitational action functional in the Newtonian limit
\label{sec:5}}

As noted above, the gravitational action functional of the TDR
contains the interaction term involving the graviton mass $m^{\vphu}_{\mbsu{g}}$.
It is known that in theories with a massive graviton based on the GR, there exists a
problem of continuity of the transition to a massless theory,
see Refs. \cite{vanDam70,Zakharov70},
as well as \cite{Rubakov08,Hinterbichler}, where this problem is discussed in detail,
and the papers \cite{Vainshtein72,Deffayet02,Babichev09,Babichev10} which consider,
including numerical modelling examples, possible solutions to this problem.
In what follows, taking into account the results obtained in
\cite{Vainshtein72,Deffayet02,Babichev09,Babichev10},
it will be assumed that for sufficiently small values of
$m^{\vphu}_{\mbsu{g}}$, the influence of the graviton mass on the character
of the interaction between particles of ordinary and dual matter is small.
This assumption and the fact that the values of $m^{\vphu}_{\mbsu{g}}$ and $\zeta$
obtained in Ref.~\cite{TDR} are extremely small,
allow one to set $m^{\vphu}_{\mbsu{g}} = 0$ and $\zeta = 0$ when deriving
the Newtonian limit of the gravitational action functional of the TDR.
In this case, using Eqs. (\ref{def:g0}), (\ref{def:hmn})--(\ref{def:h4kkl}),
and confining ourselves to the terms quadratic in $h_{\dbss{\mu\nu}}$,
the gravitational action functional (\ref{actgred})
in the Newtonian approximation can be represented as:
\be
\bar{S}^{\mbsu{NL}}_{\mbsu{g}} = S^{\mbsu{NL}}_{\vphi} + S^{\mbsu{NL}}_{\xi,\psi}\,,
\label{def:sgnl}
\ee
where
\be
S^{\mbsu{NL}}_{\vphi} = -\frac{1}{8 \pi G_{\mbsu{N}}}
\int^{\,\infty}_{-\infty} dt \int d^{\,3} x\,(\nabla\vphi(\bfx))^2\,,
\label{def:sgnlphi}
\ee
and the functional $S^{\mbsu{NL}}_{\xi,\psi}$ depends (quadratically)
only on the fields $\xi_{\dbss{k}}$ and $\psi_{\dbss{kl}}$.
The total action functional of the system ``matter plus gravitation'' in the Newtonian limit
takes the form:
\be
S^{\mbsu{NL}} = S^{\mbsu{NL}}_{\mbsu{m}\sssp} + \bar{S}^{\mbsu{NL}}_{\mbsu{g}},
\label{def:smgtotnl}
\ee
where the functional $S^{\mbsu{NL}}_{\mbsu{m}\sssp}$ is determined in
(\ref{def:smtotnl})--(\ref{def:mtot}).

%sec.6
\section{Equations of motion for the gravitational field and the effective action functional
\label{sec:6}}

The equations of motion for the fields $\vphi$, $\xi$, and $\psi$ which determine
the gravitational field are obtained from the conditions:
\be
\delta S^{\mbsu{NL}}/\delta \vphi = 0\,,
\label{eomg1}
\ee
\be
\delta S^{\mbsu{NL}}/\delta \xi_{\dbss{k}} = 0\,,\quad
\delta S^{\mbsu{NL}}/\delta \psi_{\dbss{kl}} = 0\,.
\label{eomg0}
\ee
But since the functional $S^{\mbsu{NL}}_{\mbsu{m}\sssp}$ in (\ref{def:smgtotnl})
does not depend on $\xi$ and $\psi$, the solutions of the equations (\ref{eomg0}) are
\be
\xi_{\dbss{k}} = 0\,,\quad \psi_{\dbss{kl}} = 0\,.
\label{zxipsi}
\ee
On the other hand, direct substitution of the formulas
(\ref{def:smtotnl})--(\ref{def:smgtotnl}) into Eq.~(\ref{eomg1})
leads to the problem of gravitational self-interaction (see, e.g., \cite{Arnowitt08}).
To circumvent this problem, introduce an auxiliary functional
\be
V(r_0) = \sum_{\nor} m^{\mbsu{(g)}}_{\nor} \int d^{\,3}x\,
D(|\,\bfx^{\vphu}_{\nor} - \bfx\,|, r_0)\,\vphi(\bfx),
\label{def:vaux}
\ee
where
\be
D(r,r_0) = \frac{3}{4\pi} \frac{r^2_0}{\;(r^2 + r^2_0)^{5/2}}\,,
\label{def:drr0}
\ee
$r_0$ is a parameter that tends to zero in the final formulas.
One can show that
\be
\lim_{r_{\vps_{\sscs{0}}} \rightarrow\,0} D(r,r_0) = \delta^{3} (\bfr),
\label{drr0lim}
\ee
and therefore
\be
\lim_{r_{\vps_{\sscs{0}}} \rightarrow\,0} V(r_0) = V^{\vphu}_0,
\label{vauxlim}
\ee
where $V^{\vphu}_0$ is the interaction energy of the particles
of both kinds of matter determined in Eq.~(\ref{def:tvnl}).

Define the auxiliary total action functional of the system ``matter plus gravitation''
\be
S^{\mbsu{NL}}(r_0) = S^{\mbsu{NL}}_{\mbsu{m}\sssp}(r_0) + \bar{S}^{\mbsu{NL}}_{\mbsu{g}},
\label{def:stotaux}
\ee
where the functional $S^{\mbsu{NL}}_{\mbsu{m}\sssp}(r_0)$ is determined by
Eqs. (\ref{def:smtotnl})--(\ref{def:mtot}) in which $V^{\vphu}_0$
is replaced by $V(r_0)$.
Equation for the field $\vphi$ is obtained from the condition
\be
\delta S^{\mbsu{NL}}(r_0)/\delta \vphi = 0\,.
\label{eomg1a}
\ee
It has the form
\be
\Delta \vphi(\bfx) = 4 \pi G_{\mbsu{N}}
 \sum_{\nor} m^{\mbsu{(g)}}_{\nor}
D(|\,\bfx - \bfx^{\vphu}_{\nor}|, r_0)\,.
\label{phieq}
\ee
The solution of Eq.~(\ref{phieq}), that goes to zero at infinity, is
\be
\vphi(\bfx) = - G_{\mbsu{N}}
 \sum_{\nor} m^{\mbsu{(g)}}_{\nor}
F(|\,\bfx - \bfx^{\vphu}_{\nor}|, r_0)\,,
\label{phisol}
\ee
where
\be
F(r, r_0) = 1/\sqrt{r^2 + r^2_0}\,.
\label{def:fr0}
\ee
Note that for the integral
\be
J(a, r_0) = \int d^{\,3}r\,D(r,r_0)\,F(|\,\bfr - \bfa\,|, r_0)\,,
\label{def:jar0}
\ee
where $a = |\bfa|$, the following relations are valid
\bea
J(a, r_0) &=& j(r_0/a)/a\,,\qquad
\lim_{\ve \rightarrow\,0} j(\ve) = 1\,,
\label{jar0lim}\\
J(0, r_0) &=& \frac{3\pi}{16 |r_0|}\,.
\label{jar00}
\eea

Now, define the reduced total action functional of the system ``matter plus gravitation''
by the formula
\be
S^{\mbsu{NL}}_{\mbsu{red}}(r_0) = S^{\mbsu{NL}}(r_0) +
\int^{\,\infty}_{-\infty} dt\,M^{\mbsu{eff}}_{\mbsu{g}\sssp}(r_0)c^2,
\label{def:stotred}
\ee
where the functional $S^{\mbsu{NL}}(r_0)$ is given by Eq.~(\ref{def:stotaux}),
the total effective mass $M^{\mbsu{eff}}_{\mbsu{g}\sssp}(r_0)$
is determined by the formula
\be
M^{\mbsu{eff}}_{\mbsu{g}\sssp}(r_0) = \sum_{\nor} |m^{\mbsu{(g)}}_{\nor}|
\left( 1 - \frac{3\pi G_{\mbsu{N}} |m^{\mbsu{(g)}}_{\nor}|}{32 c^2 |r_0|} \right).
\label{def:meff}
\ee
Substitution of the solution (\ref{phisol}) for the field $\vphi$ into
the r.h.s. of Eq.~(\ref{def:stotred}) gives an effective functional
$S^{\,\mbsu{eff}}(r_0)$ which depends only
on the coordinates and velocities of the particles of ordinary and dual matter.
It has the form
\be
S^{\,\mbsu{eff}}(r_0) = \int^{\,\infty}_{-\infty} dt\,
\Bigl(T - V^{\,\mbsu{eff}}(r_0)\Bigr),
\label{sltoteff}
\ee
\be
V^{\,\mbsu{eff}}(r_0) =
- \frac{G_{\mbsu{N}}}{2} \sum_{\nor \neq \norp}
\frac{j(r_0/|\bfx^{\vphu}_{\nor\norp}|)\,
 m^{\mbsu{(g)}}_{\nor} m^{\mbsu{(g)}}_{\norp}}{|\bfx^{\vphu}_{\nor\norp}|}\,,
\label{def:veff}
\ee
where $\bfx^{\vphu}_{\nor\norp} = \bfx^{\vphu}_{\nor} - \bfx^{\vphu}_{\norp}$,
the function $j(\ve)$ is determined by Eqs. (\ref{def:jar0}) and (\ref{jar0lim}),
the total kinetic energy of particles $T$ is determined in Eq.~(\ref{def:tvnl}).
The terms with $\nor = \norp$ excluded from the sum in (\ref{def:veff}) are included
in the total effective mass $M^{\mbsu{eff}}_{\mbsu{g}\sssp}(r_0)$
(with the use of formula (\ref{jar00}) in Eq.~(\ref{def:meff})).
Thereby, the effects of gravitational self-interaction are excluded from
the effective action of the particle system $S^{\,\mbsu{eff}}(r_0)$.
The contribution of these effects (which lead to the divergence
$\sim 1/r_0$ at $r_0 \rightarrow\,0$, see also \cite{Arnowitt08})
is transferred to the total effective mass (\ref{def:meff}),
which does not depend on the coordinates and velocities of the particles
and therefore (remaining finite at finite $r_0 \neq 0$)
does not affect their equations of motion.
Passing to the limit $r_0 \rightarrow\,0$ in Eqs.
(\ref{sltoteff}) and (\ref{def:veff}), one obtains taking into account (\ref{jar0lim}):
\be
V^{\,\mbsu{eff}}_0 =
\lim_{r_{\vps_{\sscs{0}}} \rightarrow\,0}
V^{\,\mbsu{eff}}(r_0) =
- \frac{G_{\mbsu{N}}}{2} \sum_{\nor \neq \norp}
\frac{m^{\mbsu{(g)}}_{\nor} m^{\mbsu{(g)}}_{\norp}}
{|\bfx^{\vphu}_{\nor\norp}|}\,.
\label{def:veff0}
\ee

This expression has the form of the gravitational interaction energy
of a system of massive particles in Newtonian mechanics. Its difference from the classical
formula lies in the replacement of the inertial masses of the particles with gravitational
ones, and this replacement explains the meaning of the term ``gravitational mass''.
For particles of ordinary matter, these masses, in accordance with the EP,
coincide. But for particles of dual matter,
$m^{\mbsu{(i)}}_{\ntld} > 0$, $m^{\mbsu{(g)}}_{\ntld} < 0$, and in the general case,
see Eq.~(\ref{mtig}), $|m^{\mbsu{(g)}}_{\ntld}| \neq |m^{\mbsu{(i)}}_{\ntld}|$.

Note, however, that the functional $S^{\,\mbsu{eff}}(r_0)$, Eq.~(\ref{sltoteff}),
is invariant under the transformations
\begin{subequations}
\bea
&&\bfx = \alpha\,\bfx', \qquad t = \beta\,t',
\label{def:xttrans}\\
&&m^{\mbsu{(i)}}_{\nor} = \frac{\beta}{\alpha^2}\,m^{\prime\mbsu{(i)}}_{\nor},\qquad
m^{\mbsu{(g)}}_{\nor} = - \beta^{-1}\,m^{\prime\mbsu{(g)}}_{\nor},
\label{def:migtrans}\\
&&G^{\vphu}_{\mbsu{N}} = \alpha\beta\,G^{\prime}_{\mbsu{N}},
\qquad r^{\vphu}_0   = \alpha r'_0,
\label{def:gntrans}
\eea
\end{subequations}
where $\alpha$ and $\beta$ are the constants.
At $\alpha = A^2$ and $\beta = B^2$, these transformations reduce
the metric $\gtld^{(0)}_{\dbss{\mu\nu}}$ to the form (\ref{def:g0}),
and lead to the relations
\bea
&&
m^{\prime\mbsu{(i)}}_{n} = \frac{A^4}{B^2}\,m^{\vphu}_{n},\qquad
m^{\prime\mbsu{(g)}}_{n} = - \frac{B^4}{A^4}\,m^{\prime\mbsu{(i)}}_{n},
\label{mpign}\\
&&
m^{\prime\mbsu{(i)}}_{\ntld} = m^{\prime\mbsu{(g)}}_{\ntld} = m^{\vphu}_{\ntld}\,,
\label{mpignt}
\eea
instead of Eqs. (\ref{mig}) and (\ref{mtig}).
Thus, the equality of inertial and gravitational masses of the particles of dual matter
can be obtained, but at the price of violating this equality for
the particles of ordinary matter.
This fact is an illustration of the principle of alternative equivalences
formulated in Sec.~\ref{sec:2}.

Repeating the reasonings used in the Introduction in relation to Eq.~(\ref{def:v12}),
one concludes that the interaction between particles
belonging to the same kind of matter has the character of gravity
(however, with an adjustment for the inequality of the inertial and gravitational
masses of the particles of either dual or ordinary matter).
The interaction between particles belonging to different kinds of matter
has the character of antigravity.

%sec.7
\section{Conclusions}

In the paper, the antigravity effects arising in the theory of dual relativity (TDR)
are analyzed.
The general framework of the TDR is described, and some results obtained previously
in the cosmological applications of this theory are summarized.
In particular, quality of description of the available data on the dependence
of the Hubble parameter $H$ on the redshift $z$ obtained in the TDR is better than
in the $\Lambda$CDM model which is the standard model of modern cosmology.
At the same time, the TDR is free of the known cosmological problem of
the initial singularity and of the cosmological constant problem.

The full action functional of the TDR includes the action functionals of matter fields
of two kinds: ordinary and dual.
In the main part of the paper, the case of a system of point-like massive particles
is considered. It is supposed that these particles interact
with each other only by means of the forces determined by the properties of the metrics
that govern their motion.
In this case, the action functional of ordinary matter depends only on the metric $g$,
while the action functional of dual matter depends only on the metric $\tilde{g}$,
related to $g$ by the formula $\tilde{g} = \gamma\,g^{-1}\gamma$, in which $\gamma$
is a background flat metric.
Formulas are obtained for the effective action functional of such a system
in the Newtonian limit.
This functional includes an interaction term that has the form of the energy
of gravitational interaction in Newtonian mechanics.
However, unlike the classical formula, it contains
gravitational masses of particles instead of inertial masses.
The gravitational masses of particles of ordinary matter are positive and equal
to their inertial masses.
The gravitational masses of particles of dual matter are negative
(while their inertial masses are positive) and generally differ from the inertial masses
in absolute magnitude.
This fact signifies a violation of the conventional equivalence principle,
which implies the equality of inertial and gravitational masses
of all the bodies and particles.
It is, however, shown, that using the simple transformations of the quantities which enter
the effective action functional, one can invert these relationships between the masses.
These transformations result in
the equality of inertial and gravitational masses of the particles of dual matter,
but at the price of violating this equality for the particles of ordinary matter.
This is an illustration of the principle of alternative equivalences formulated in the paper.
In any case, the gravitational masses of the particles of ordinary and dual matter
have the opposite signs, and this difference in the signs
leads to repulsion between these particles.
This difference is a consequence of the above-mentioned relationship between the metrics
$g$ and $\tilde{g}$, which thereby underlies the antigravity mechanism in the TDR.

Certainly, realization of this mechanism is possible only if the dual matter exists,
what is currently an open question.
One of the general arguments in favor of its existence is that
the solutions of the TDR equations in the cosmological limit include dual matter as
a necessary element of a theoretical scheme compatible with the concept of a stable Universe,
the equilibrium metric of which coincides with the background flat metric $\gamma$.
Another, empirical argument is connected to the cosmic voids, which are observed
in the visible part of the Universe, and the formation of which could perhaps
be attributed to the presence of antigravitating dual matter.

\bibliographystyle{apsrev4-2}
\end{document}